\documentclass[aps, prd, a4paper, 10pt, showpacs, superscriptaddress, floatfix, nofootinbib, reprint]{revtex4-1}
\usepackage{color}
\usepackage{xcolor}
\usepackage{amsmath,amssymb,graphicx}
\usepackage{wasysym}
\usepackage{bm,lipsum}
\usepackage{times}
\usepackage{microtype}
\usepackage[utf8]{inputenc}
\usepackage{booktabs}
\usepackage{subfigure}
\usepackage[normalem]{ulem}
\usepackage[varg]{txfonts}
\usepackage{bm,graphics,graphicx,epsfig,color,ulem}
\usepackage[colorlinks, pdfborder={0 0 0}]{hyperref}
\definecolor{LinkColor}{rgb}{0.75, 0, 0}
\definecolor{CiteColor}{rgb}{0, 0.5, 0.5}
\definecolor{UrlColor}{rgb}{0, 0, 0.75}
\hypersetup{linkcolor=LinkColor}
\hypersetup{citecolor=CiteColor}
\hypersetup{urlcolor=UrlColor}
\allowdisplaybreaks
\begin{document}

%%%%%%%%%%%%%%%%%%%%%%%%%%%%%%%%%%%%%%%%%%%%%%%%%%%%%%%%%%%%%%%%%%%%%%%%%%%%%%%%%%%%%%%%%%%%%% 
%-----------------------------------------TITLE----------------------------------------------%
%%%%%%%%%%%%%%%%%%%%%%%%%%%%%%%%%%%%%%%%%%%%%%%%%%%%%%%%%%%%%%%%%%%%%%%%%%%%%%%%%%%%%%%%%%%%%% 

\title{Multiparameter tests of general relativity using multiband gravitational-wave observations}
\author{Anuradha Gupta} \email{agupta1@olemiss.edu}
\affiliation{Department of Physics and Astronomy, The University of Mississippi, Oxford MS 38677, USA}
\affiliation{Institute for Gravitation and the Cosmos, Department of Physics, Penn State University, 
University Park PA 16802, USA}
\author{Sayantani Datta} %\email{sdatta@cmi.ac.in} 
\affiliation{Chennai Mathematical Institute, Siruseri, 603103, India}
\author{Shilpa Kastha}% \email{shilpakastha@imsc.res.in}
\affiliation{Max Planck Institute for Gravitational Physics (Albert Einstein Institute), 
Callinstra\ss e 38, D--30167 Hannover, Germany \\ Leibniz Universit\" at Hannover, D--30167 Hannover, Germany \\}
\author{Ssohrab Borhanian}
\affiliation{Institute for Gravitation and the Cosmos, Department of Physics, Penn State University, 
University Park PA 16802, USA}
\author{K. G. Arun} % \email{kgarun@cmi.ac.in} 
\affiliation{Chennai Mathematical Institute, Siruseri, 603103, India}
\affiliation{Institute for Gravitation and the Cosmos, Department of Physics, Penn State University, 
University Park PA 16802, USA}
\author{B. S. Sathyaprakash} % \email{bss25@psu.edu} 
\affiliation{Institute for Gravitation and the Cosmos, Department of Physics, Penn State University, 
University Park PA 16802, USA}
\affiliation{Department of Astronomy and Astrophysics, Penn State University, University Park PA 16802, USA}
\affiliation{School of Physics and Astronomy, Cardiff University, Cardiff, CF24 3AA, United Kingdom}
\date{\today}

\begin{abstract}
In this Letter we show that multiband observations of stellar-mass binary black holes
by the next generation of ground-based observatories (3G)  and the
space-based Laser Interferometer Space Antenna (LISA) would facilitate a
	comprehensive test of general relativity by simultaneously measuring  all the
post-Newtonian (PN) coefficients. Multiband observations would measure most of
the known PN phasing coefficients to an accuracy below a few percent---two
	orders-of-magnitude better than the best bounds achievable from 
	even `golden' binaries in the 3G or LISA bands.
Such multiparameter bounds would play a pivotal role in constraining the
parameter space of modified theories of gravity beyond general relativity.
\end{abstract}

\maketitle

\paragraph*{Introduction:}
Gravitational wave (GW) observations have provided a first glimpse of the
strong-field dynamics of binary black holes (BBHs) \cite{Discovery,GWTC1}. They
have also allowed us to place the first ever constraints on the possible departures from
general relativity (GR) \cite{TOG,GWTC-TGR} in this regime. Parametrized tests
of the  post-Newtonian (PN) approximation to GR \cite{AIQS06b, MAIS10,
YunesPretorius09, LiEtal2011} are among the most important theory-agnostic, null-tests
of GR that are performed using GW observations.  These tests make use of the
analytical prediction of the structure of the phase evolution using the PN
approximation to GR~\cite{Bliving}.  In the PN approximation the dynamics of the
binary is treated as an adiabatic process and Einstein's field equations are
solved under the assumption of \emph{slow motion} and \emph{weak gravitational
fields}.  This is an excellent approximation for the ``inspiral" phase of the
compact binary dynamics where the two stars  spiral-in under the influence of
radiation back reaction, but the time-scale of radiation reaction is large
compared to the orbital time-scale.

Gravitational waveform from a compact binary coalescence, in the frequency domain, have  
the well-known form \cite{Sathyaprakash:1991mt}
\begin{equation}
{\tilde h}(f)={\cal A} f^{-7/6} e^{i\Phi (f)}\,,
\end{equation}
where $\Phi(f)$ is the frequency domain phase of the emitted signal and ${\cal
A}$ is the signal's amplitude. For inspiralling binaries in quasi-circular orbits,  the waveform
depends on the binary's masses, spins, distance, sky position, and the
orientation of its orbit. More explicitly, the phase takes the form
\begin{equation}
\Phi(f)=2\pi f\,t_c-\phi_c+\frac{3}{128\,\eta\,v^5}\left[ \sum_{k=0}^K
\phi_k\,
v^k+\sum_{kl=0}^K\phi_{kl}\,v^{kl}\ln v\right]\,,
\end{equation}
where $v=(\pi m f)^{1/3}$ denotes the PN expansion parameter,  $m$ denotes the binary's 
total mass, and  $\phi_{kl}$ and $\phi_k$ denote the logarithmic and
non-logarithmic phasing coefficients, respectively. The PN coefficients are currently
known up to 3.5 order in the PN expansion \cite{BDIWW95,BDEI04,ABFO08,MKAF16}, 
which corresponds to $k=7$ in the above equation.   Various PN coefficients
capture a range of nonlinear interactions and physical effects in GR~\cite{Bliving}.
These include
the effect of mass asymmetry (1PN and above), 
different types of `tail' effects (1.5PN, 2.5PN, 3PN, 3.5PN)~\cite{BD87}  as well as 
physical interactions such as spin-orbit (1.5PN, 2.5PN, 3PN, 3.5PN)~\cite{KWWi93,BBuF06} 
and spin-spin effects (2PN, 3PN)~\cite{KWWi93,BFH2012}, and effects due to the 
presence or absense of a horizon of the compact objects (2.5PN)~\cite{Tagoshi:1997jy}. 
(See Ref.~\cite{YYP2016} for an in-depth discussion about the modifications to the GR phasing formula from various modified theories).
The parameters $t_c$ and
$\phi_c$ are the epoch when the signal's amplitude at the detector is the greatest 
and the phase of the signal at that epoch, respectively.  For BBHs on quasi-circular 
orbits, the PN coefficients $\phi_k$ and $\phi_{kl}$ are functions of the component 
masses and spins.  The assumption of a quasi-circular orbit is an excellent 
approximation for majority of the stellar-mass BBHs \cite{Buonanno:2009zt}.

The parametrized tests rely on the unique prediction for the
PN coefficients $\phi_k$ and $\phi_{kl}$ in GR and use GW BBH
merger events to constrain possible departures of the coefficients from 
their GR prediction. A parametrized waveform replacing the GR phasing coefficients 
$\phi_a$ with $\phi_a (1+\delta{\hat \phi}_a)$ ($a=k, kl$) is employed 
for the test~\cite{LiEtal2011}. By construction,  the
\emph{deformation} parameters $\delta{\hat \phi}_a$ are identically equal to zero
in GR, while in  a modified theory of gravity one or more of these
parameters  can deviate from zero. Thus, GW data allow
the direct measurement of the PN coefficients  and if their deformations are found to be
consistent with zero, the uncertainty associated with the measurement
provides an upper limit on the deviation of these parameters from their GR values.

\paragraph*{Status of parametrized tests of post-Newtonian theory:}
Combining data for the ten BBH merger events found during the first and second
observing runs of LIGO/Virgo, the current bound on the eight PN deformation parameters are
given in Fig.\,4 of Ref.\,\cite{GWTC-TGR}. Moreover, the bounds from this
theory-agnostic test have been mapped onto specific modified theories of
gravity in Ref.\,\cite{YYP2016}.  However, there is an important caveat while
using these bounds to constrain a modified theory of gravity: The bound
on the deviation from a particular PN coefficient reported in
Ref.\,\cite{GWTC-TGR} is derived assuming that \emph{all} the deformation parameters 
{\it except} the one that is being tested follow the predictions of GR with
$\delta{\hat { \phi_a}}=0$. This assumption is necessary because the most
general test wherein all the PN coefficients are simultaneously measured yields
very poor or no bounds due to the strong degree of covariance among the
deformation parameters and the intrinsic parameters of the binary~\cite{AIQS06a}.
Hence one is compelled to replace this most general test with a series of tests
wherein only one deformation parameter is varied at a time together with, of
course, the intrinsic parameters of the binary. This restricted suite of tests
can still be expected to detect a deviation away from GR by finding
statistically significant offsets away from zero in one or more of the PN 
 deformation parameters~\cite{MAIS10, LiEtal2011}. 

 It would, however, be misleading to use the results from the single-parameter tests to constrain
a specific physical effect in a modified theory of gravity. Firstly, any
deviation from GR inferred for a particular PN coefficient cannot be attributed
{\it uniquely} to a breakdown of GR at that PN order. This is because the
waveform is largely degenerate in the PN coefficients. Consequently, deviation
at a particular PN order can be captured by deformation of the coefficient at a
different PN order. Hence a deviation in one or more of the PN coefficients in
a set of tests does not necessarily give any fundamental insight into the true
nature of the underlying theory of gravity.  Secondly, if the single-parameter
tests are all consistent with GR, the widths of the posterior distributions of
the PN coefficients cannot be used to constrain the parameter space of modified
theories of gravity.  This is
due to the expectation from effective field theoretic arguments, that
deviations from GR, in a specific modified theory of gravity, show up starting
\emph{from} a certain PN order (see, for instance, \cite{Endlich:2017tqa,Sennett:2019bpc}).
Therefore, to map the PN deformation parameters to the free parameters
of a specific modified theory of gravity it is necessary to perform the most
general multiparameter test.  In other words, single-parameter tests would lead
to an underestimation of the errors and hence yield bounds that are more
stringent than what one might infer with multiparameter tests.

In this {\it Letter} we will show that combining data from the next 
generation (3G) of ground-based detectors, such as the Cosmic Explorer (CE)
\cite{2019arXiv190704833R} and Einstein Telescope (ET) \cite{Punturo:2010zz}, 
with the space-based LISA observatory \cite{LISA2017} is likely the only viable 
route to carry out this very challenging, but very general test of GR. Such tests 
are crucial to set reliable constraints on the parameter space of modified theories 
of gravity.  Specifically, we demonstrate that multiband observations 
of a subclass of stellar-mass BBHs by LISA and CE would provide a unique opportunity
to carry out the multiparameter test of PN theory. Combining the low-frequency 
sensitivity of LISA with the high-frequency sensitivity of CE helps in lifting 
the large degeneracies that prevent the use of multiparameter tests in either 
of these observatories. To demonstrate the advantage of multiparameter tests 
using multibanding we simulate a stellar-mass population of BBHs that obey the 
mass-, rate- and redshift-distribution inferred from the first and second observing 
runs of LIGO and Virgo.  In a companion paper \cite{Datta:2020vcj},  we
will discuss intermediate-mass BBHs as another important class of
sources for multiband, multiparameter test of GR, although the bounds 
from stellar-mass BBHs are far better than their intermediate-mass 
counterparts~\cite{Datta:2020vcj}.

\paragraph*{Multiband visibility of stellar-mass binaries:}
The planned LISA observatory is sensitive to GWs in the frequency range $\sim $
0.1--100 mHz and the proposed  3G observatories (e.g., CE, which we have
used in this paper as a representative of 3G detectors), will be sensitive in
the frequency range $\sim $ 1 Hz -- 5 kHz.  Though LISA is more sensitive to
mergers of supermassive BBHs of millions of solar masses, it has been argued
that the detection of stellar-mass BBHs using LISA would be possible despite
the small signal-to-noise ratios
(SNRs)~\cite{Nair:2015bga,Vitale:2016rfr,Sesana:2016ljz} and would be of
immense importance to astronomy and fundamental physics, as the mergers of
these binaries would be detectable by the ground-based detectors operating at
the same time. Observation of sources at earlier stages of their evolution in
LISA, and later, more nonlinear, stages in 3G detectors is referred to as
\emph{multiband} observation.

Several authors have investigated the value added by multiband observations of
GW sources.  For example, Ref.\,\cite{Barausse:2016eii,Toubiana:2020vtf,Liu:2020nwz} examined the projected
constraints on the bounds on dipolar GW radiation, Ref.\,\cite{Carson:2019rda}
investigated the bounds on single-parameter tests of GR and
Ref.\,\cite{Gnocchi:2019jzp} studied the constraints on the parameter space of
modified theories of gravity using multiband observations. These authors have
used prototypical BBH systems, such as GW150914 \cite{Discovery},  which will
have good multiband visibility, and have studied the corresponding bounds for
single-parameter tests of GR.

Here, we consider $5\times10^5$ BBHs corresponding to one year of CE observation
\cite{Baibhav:2019gxm}, distributed uniformly in comoving volume up
to redshift $z=10.$ The primary black hole masses are assumed to follow a
power-law distribution with the power-law index $\alpha=1.6$ 
(i.e., $p(m_1)\propto m_1^{-\alpha}$) in the mass range
$[5, 100]M_{\odot}$  while secondary masses are uniform in the same mass range~\cite{LIGOScientific:2018jsj}. 
We assume the binary components to possess spins which are aligned or anti-aligned 
with respect to the orbital angular momentum vector.  This assumption is consistent with the fact that none of the
BBHs detected during the first and second observing runs of LIGO/Virgo showed
evidence for spins misaligned with the binary's orbital angular momentum.
The Kerr parameter of the companion black holes are drawn from two
different distributions: (1) a uniform distribution in the range $[0,1]$ and (2)
a Gaussian with mean 0 and standard deviation 0.1.  

\begin{figure}
\includegraphics[width=0.5\textwidth]{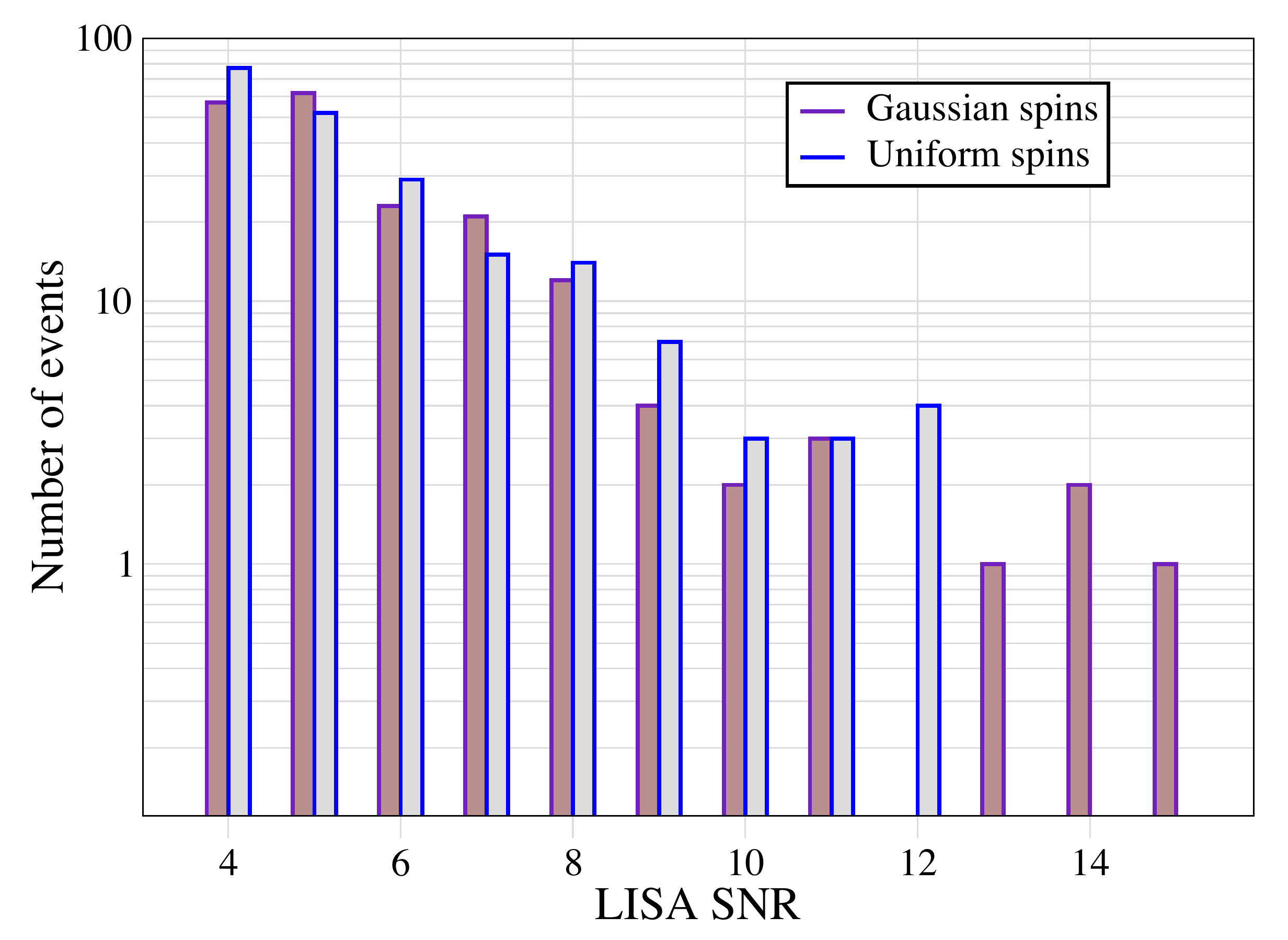}
\caption{
Distribution of the SNR of stellar-mass 
BBHs in the LISA band using the mass and redshift distribution as
inferred from the first and second observing runs, and spins following 
(i) a uniform distribution and (ii) a Gaussian distribution with mean 0 and 
standard deviation 0.1. Only $\rm SNR\ge4$ events are shown. A small fraction of 
about  ${\sim}200$  of all sources 
(some 500,000) observed by CE in a year have SNR $\geq4$ in the LISA band.
The plot also shows that the spin distribution of black holes
does not have a significant effect on the visibility of stellar-mass BBHs in LISA.}
\label{fig:LISA-SNR}
\end{figure}

3G detectors will be able to observe stellar-mass BBH mergers up to the epoch 
of the formation of first stars.  The question is what fraction
of events detected by CE will have LISA counterparts. This joint
population will be limited by the SNR in the LISA band.  In Fig.~\ref{fig:LISA-SNR}
we plot the SNR distribution  in LISA for this population. As expected, only
a small subset of the population will have an SNR greater than 4.
Such events will have an SNR of at least 2000 in CE, facilitating 
a very accurate measurement of the binary parameters, which in turn helps in
digging the signals out of the LISA background noise. For example, the chirpmass 
and the symmetric mass ratio are both measured to an accuracy better than a few 
parts in a million and the source's position on the sky will be determined to within
75 square arcminutes (enough to identify the host galaxy within 500 Mpc). Consequently,
the number of templates required in archival searches of the LISA data is 
$< \rm few \times 10^4$ for 90\% of the events, orders of magnitude smaller than 
previous estimates \cite{Wong:2018uwb}. This reduces the false alarm probability of
the archival search and makes it possible to identify signals of ${\rm SNR} > 4$ 
in the LISA data with a false alarm probability of $<10^{-3}$.
(See the Supplement for 
a discussion on the detectability of stellar-mass BBH signals in LISA with 
an SNR threshold as low as 4.) 
We find that among the hundreds of thousands of 
stellar-mass BBH merger that would be observable by CE in one year, $\sim$200 
would cross this threshold of $\rm SNR>4$ and permit multiband, multiparameter tests of GR.
These $\sim$200 BBHs would spend roughly 4.5 days to 7 weeks outside the LISA 
band before entering the CE band and eventually merge. 
% \ag{We can add some more sentences about detectability around this para.} 

\paragraph*{Multiparameter tests of GR via multiband GW observations:}
We now describe the efficacy of the multiparameter tests of GR
using the population of ${\sim}200$ BBH merger events detectable by both CE and
LISA. Our method here is based on the well-known Fisher information
matrix which enables the computation of the projected $1\sigma$ errors on the
various parameters describing a signal model for a given sensitivity of
the detector configuration~\cite{Rao45,Cramer46,CF94}. We use the sensitivity
curves of CE and LISA given in \cite{2019arXiv190704833R} and \cite{LISA2017},
respectively.  For simplicity, we do not consider the orbital motion of
LISA as it is likely to have negligible impact on the parameter
estimation of the intrinsic parameters of the binary, which is of
interest here. Stellar-mass BBHs that merge in the CE band are
assumed to have been observed for five  years in LISA and the starting 
frequency for the signal in LISA is chosen accordingly 
following the prescription given in Eq.~(2.15) of Ref.\,\cite{BBW05a}.

We employ the IMRPhenomD~\cite{Husa2016, Khan2016} waveform model, a frequency-domain 
phenomenological model describing the complete inspiral-merger-ringdown phases 
of BBH systems. The waveform amplitude in this model is truncated at the quadrupolar 
order and we have introduced additional deformation parameters $\delta\hat\phi_a$
in the phase at different PN orders in the inspiral part of the waveform.  We have set 
the four angles corresponding to the sky position of the binary and the orientation of
its orbit  with respect to the line-of-sight to zero. This amounts to assuming that the binaries are optimally located
and oriented with respect to the detectors. Note, however, that the LISA 
sensitivity curve that we use is averaged over the sky and the polarization angle 
and we have included a factor of $\sqrt{4/5}$ in the calculation of the SNR and the Fisher matrix
to account for the  averaging over the inclination angle \cite{Cornish:2018dyw}.

The Fisher information matrix for a single detector (CE or LISA) is defined as
\begin{equation}
\label{eq:fisher}
\Gamma_{\alpha\beta}^{(0)}=\langle\tilde{h}_\alpha,\tilde{h}_\beta\rangle,
\end{equation}
where $\tilde{h}(f;\vec{\theta})$ is the GW signal defined 
by a set of parameters ${\vec \theta}$, $\tilde{h}_\alpha=\partial\tilde{h}(f;\vec{\theta})/\partial\theta_\alpha$, 
and the angular bracket $\langle\,,\, \rangle$ denotes the noise-weighted inner product 
defined by 
\begin{align}
\label{eq:innerproduct}
\langle a,b\rangle=2\int_{f_{\rm low}}^{\rm f_{\rm high}}\frac{a(f)\,b^*(f)+a^*(f)\, b(f)}{S_h(f)}\,df \,,
\end{align}
where $S_h(f)$ is the one-sided noise power spectral density (PSD) of
the detector and $f_{\rm low}, f_{\rm high}$  are the lower and upper
limits of integration. For CE the lower limit of integration is taken to be 5 Hz and
the upper frequency cut-off is chosen such that the characteristic amplitude ($2 {\sqrt f} |\tilde h(f)|$)
of the GW signal is lower than that of the CE noise by 10\% at maximum. 

In order to combine the information from LISA and CE, we construct a
multiband Fisher matrix by simply adding the Fisher matrices for the
individual detectors, with the corresponding variance-covariance matrix $C^{\alpha\beta}$
defined by the inverse of the multiband Fisher matrix: 
\begin{equation}
\label{CombFish}
\Gamma_{\alpha\beta}=\Gamma_{\alpha\beta}^{\text{CE}}+\Gamma_{\alpha\beta}^{\text{LISA}},
\quad C^{\alpha\beta}=(\Gamma^{-1})^{\alpha\beta},
\end{equation}
The diagonal components, $C^{\alpha\alpha}$, are the variances of $\theta^\alpha$ and
the 1$\sigma$ errors on $\theta^\alpha$ are $\sigma^\alpha = \sqrt{C^{\alpha\alpha}}.$

The errors $\sigma_a$, where $a=1,2,\cdots, 8$ denote the deformation parameters
that are tested simultaneously, are obtained for each event in the population for 
different choices of the number of test parameters $\delta\hat\phi_a,$ $a=1, 2, 
\cdots, 8.$  The bounds on the individual events are combined to obtain a net 
constraint by using the standard formula
\begin{equation}
{\sigma_a^{-2}}=\sum_{n=1}^{N}
{\left(\sigma_a^{(n)}\right)^{-2}},
\end{equation}
where $n=1,\ldots,N$ denotes the events in the BBH population and $N$ is their total number.

Following Refs.~\cite{CF94,PW95} we also add a prior matrix $\Gamma^p$ to the
Fisher information matrix, $\Gamma^{(0)}$, in order to account for certain
properties of the signals that we assume.  Specifically, we assume that the priors on
the spin magnitudes and the phase of coalescence as
$\Gamma^p_{\chi_1\chi_1}=\Gamma^p_{\chi_2\chi_2}=(0.5)^{-2}$ and
$\Gamma^p_{\phi_c\phi_c}=(\pi)^{-2},$ respectively and all other elements of the
prior matrix are set to zero.  The Gaussian prior on spin magnitudes is a good
approximation to the low-component spins of the BBHs reported in Ref.~\cite{GWTC1}.
The prior on $\phi_c$ is somewhat adhoc, but helps the Fisher matrix to be
better conditioned. We have verified that this choice of prior does not alter
our conclusions reported in this paper.  We now invert the resulting Fisher
matrix given by $\Gamma_{\alpha\beta}=\Gamma^0_{\alpha\beta}+\Gamma^p_{\alpha\beta}$ 
to deduce the error bars.

\begin{figure}
\includegraphics[width=0.48\textwidth]{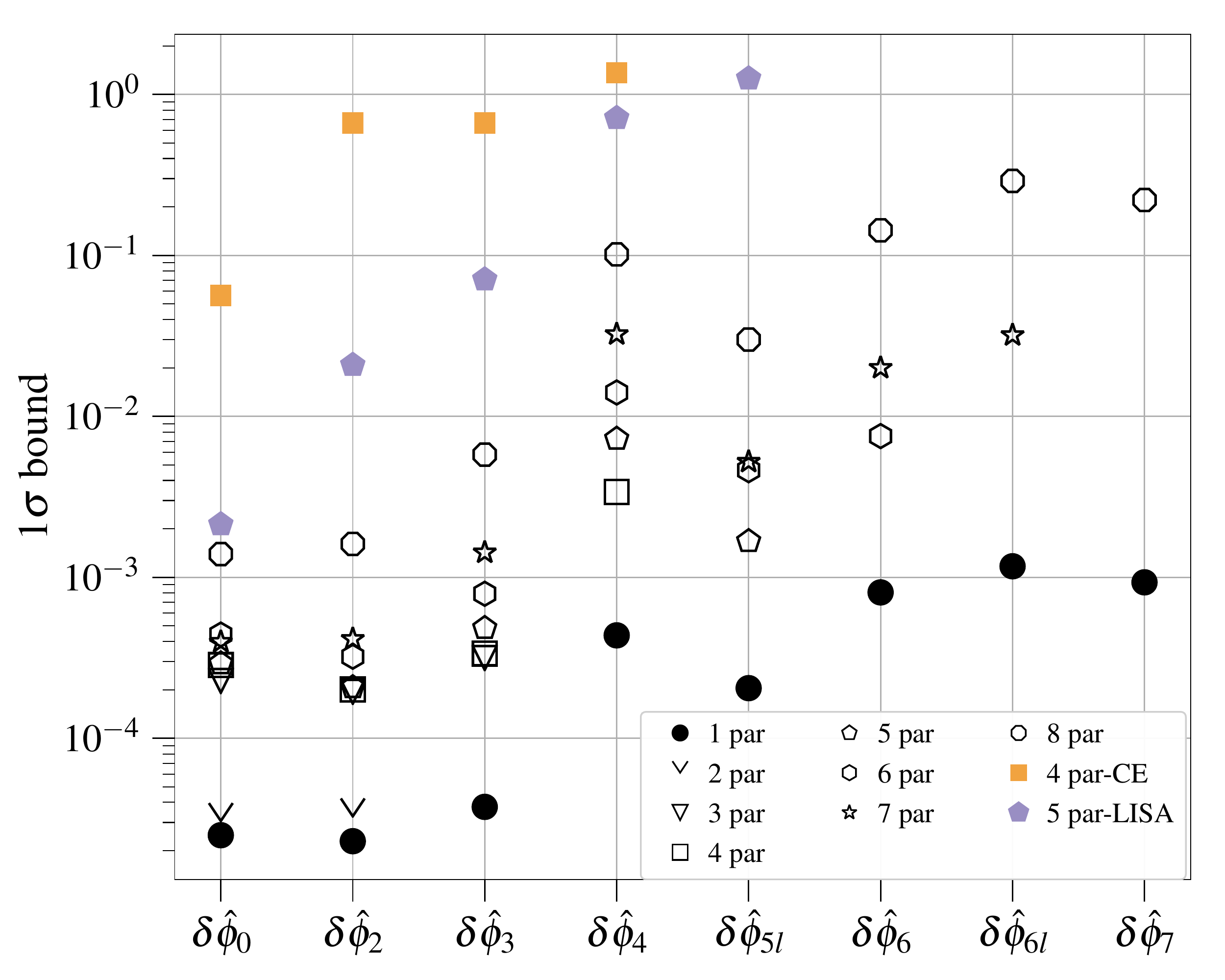}
\caption{Multiparameter tests using multiband observations with LISA and CE. 
         Shown are combined $1\sigma$ bounds on various PN coefficients starting 
         from 0PN through 3.5PN in the inspiral phase of the signal while measuring 
         many of them together at a time. Different types of markers symbolize how 
         many PN deformation parameters were constrained simultaneously. 
         For example, `$\CIRCLE$' represents `one PN deformation parameter at a time’, 
         `$\vee$' represents `two PN deformation parameters at a time’, and so on. 
         The figure represents results for the BBH population having Gaussian spin 
         distribution, we get similar estimates for a uniform spin distribution. 
         The filled diamonds and pentagons are bounds obtained with CE and LISA,
         respectively, on the first four and five PN deformation parameters from their 
         respective golden binaries, respectively.  The total masses of the CE 
         and LISA golden binaries are $200 \,M_{\odot}$ 
         and $6.6\times10^{5}\,M_{\odot}$, both binaries are 1 Gpc away and have 
         component spins $\chi_1=0.6, \chi_2=0.5$.}
         \label{fig:boundsleft}
\end{figure}

\paragraph*{Results and Discussions:}\label{sec:results}
Our main results combining LISA and CE observations of stellar-mass BBHs 
are summarized in Fig. \ref{fig:boundsleft}. 
As we increase the number of PN coefficients that are simultaneously tested, starting
from the Newtonian order,  the 1$\sigma$ upper bounds on them are presented in the 
figure.  For instance, the filled circles are the bounds where only one PN deformation 
parameter is estimated at a time, whereas the octagons denote the bounds when all
the eight parameters are simultaneously estimated. In the eight parameter case, 
all the parameters are measured with an accuracy ${\sim}20\%$, of which the first 
three may be measured with an accuracy better than 1\%, whereas the first two PN 
coefficients may yield bounds ${\sim}0.1\%$. 

One may notice interesting trends in the bounds as we add more and more
parameters.  The bounds on 0PN and 1PN deformation coefficient from 2-parameter
estimation case are $<0.01\%$. The inclusion of the 1.5PN
deformation coefficient results in a sudden worsening of the bounds by an order
of magnitude. This may be understood by noting that 1.5PN is the order at which
spins first appear in the phasing formula. Adding a deformation parameter at
that order, that is completely degenerate with spins, adversely affects the
overall parameter estimation, which gets reflected in the bounds on the first two
PN coefficients. The gradual worsening of the bounds as we go to even higher
multiparameter tests is simply due to the increasing degeneracy brought in by
each of the additional PN deformation parameters.
Nevertheless, multiband observations of
stellar-mass BBHs would permit us to test modified theories of gravity, which
predict deviations at orders below 3PN to a precision less than $1\%$.

It can be seen that even in the era of 3G detectors we cannot obtain meaningful 
constraints with multiparameter tests. As is evident from Fig.~\ref{fig:boundsleft}, for golden 
binaries in CE---binaries that have the smallest error for the multiparameter tests---
the errors on $\delta\phi_a,$ are ${\sim}100\%$  only for $a=1,\ldots, 4$; if 
we vary more than four parameters at a time then the errors on PN coefficients with $a\neq0$ 
are larger than 100\%. In a year's time CE will observe a handful of such golden binaries and  
the joint error that one can obtain by combining golden binaries will still not 
be significantly smaller. Consequently, ground-based detectors alone cannot 
break the degeneracy among different PN coefficients.  The same is true with 
LISA observations of supermassive BBHs. Even with a golden supermassive BBH we 
can perform the multiparameter test with only five parameters and LISA is not 
likely to observe more than a handful of such binaries over a five-year period.
Having said this, in this paper we do not consider other ways to compute 
the combined bounds on $\delta\phi_a,$ such as by combining all the events observed 
in CE and LISA, as our method already achieves the desired accuracy needed for the multiparameter test.

\paragraph*{Conclusions:} 
To conclude, we have shown the importance of multiband observations of GWs to
carry out the multiparameter tests of GR. From our systematic study of a
representative set of systems, we have also found that even for the best case
scenario, observations of supermassive BBHs in the LISA band or stellar-
or intermediate-mass BBHs in the CE band would not be able to place
constraints as good as the one reported here. Hence multibanding would, perhaps,
be the only way to carry out this test which in turn is necessary to make
meaningful constraints on the parameter space of modified theories of gravity.
As LIGO and Virgo detect several more BBHs in the future observing runs, the
merger rate and the mass distribution would be more tightly constrained which is
likely to further tighten the bounds derived here making this test an excellent
science case for multiband observations.

\paragraph*{Acknowledgments:} 
We thank Chris Van Den Broeck, Bala Iyer, Arnab Dhani
and M. Saleem for several useful discussions.
B.S.S. is supported in part by NSF Grant No. PHY-1836779, AST-1716394 and
AST-1708146.  S.B. is supported by NSF Grant No. PHY-1836779. 
K.G.A. and S. D. are partially supported by a grant from the Infosys
Foundation. They also acknowledge the Swarnajayanti grant
DST/SJF/PSA-01/2017-18 DST-India. K.G.A acknowledges Core Research Grant
EMR/2016/005594 of SERB. We acknowledge the use of IUCAA LDG cluster Sarathi for the computational/numerical work.
We thank all frontline
workers combating the CoVID-19 pandemic without whose support this work would not
have been possible.

\appendix*
\section{Supplemental Materials}

In this Supplement we provide a discussion of the detectability of gravitational waves
from stellar-mass binary black holes (BBHs) by the Laser Interferometer Space
Antenna (LISA), an alternative to the multiparameter test presented in the paper  and
the accuracy of Fisher matrix inversion.

\paragraph*{Archival searches for stellar-mass BBHs in the LISA data:}
Ground-based observatories such as the Cosmic Explorer (CE) \cite{2019arXiv190704833R} 
and Einstein Telescope (ET) \cite{Punturo:2010zz}, have the best sensitivity to 
stellar-mass BBHs of 5-100 $M_\odot$ and can detect them up to redshifts of $z\sim10$ 
and beyond. A small fraction of such BBHs that are close enough will also be observable 
by LISA \cite{LISA2017}. The observability of a source depends on the false alarm rate at a 
given signal-to-noise-ratio (SNR) and the number of trials needed to dig out the signal. 

As an example, there is only a chance of 0.13\% (i.e., a p-value of 0.0013)
that a single draw from a Gaussian distribution with zero mean and unit variance 
yields a number larger than 3.  On the other hand, multiple draws from the same 
distribution increases the p-value for getting a number larger than 3. 
Likewise, if it is necessary to carry out a 
blind search for GW signals without any knowledge of their parameters then one ought to 
employ a large number of templates which makes it computationally expensive 
to dig out weaker signals 
from a noisy background.  Indeed, Ref.~\cite{Moore:2019pke} argued that as many
as $10^{40}$ templates would be needed to dig out a stellar-mass BBH signal from 
LISA data. This would require a matched filter SNR $\geq14$ for a p-value of 
$10^{-3}$. 

A third-generation (3G) network of CE and ET can aid in searching for stellar-mass BBH signals in LISA 
data as the former will detect them with extremely high fidelity and can therefore provide 
a very tight prior on the source's extrinsic (sky position, polarization, orientation 
of the orbit, and luminosity distance) as well as intrinsic (masses and spins of the 
companion black holes) parameters. Therefore, an archival search for signals can greatly 
reduce the number of templates/trials required and hence enhance their detectability 
in the LISA data.

The analyses in Ref.\,\cite{Moore:2019pke} assumed for the LISA archival search 
that the trigger time, the epoch when the signal's amplitude reaches its maximum
value in the detector, is known precisely, while the errors in the intrinsic and other 
extrinsic parameters will be smaller in 3G detectors than those measured by
LIGO \cite{Discovery} by a factor of 10. This contracted volume of the search space
decreases the number of templates by a factor of $10^{29}$ compared to 
a blind search \cite{Moore:2019pke}.  Yet the result is that LISA would need 
$\sim10^{11}$ templates to identify a stellar-mass BBH signal, or an SNR 
threshold of $\sim$9 for a p-value of $10^{-3}.$ Other studies \cite{Wong:2018uwb} 
have shown that using an alternate method one can detect GW150914-like 
source even with an SNR$\sim$7 for the same p-value.  We have computed the 
number of templates for the events considered in this work and find that LISA 
archival searches require far fewer templates owing to the greatly contracted
search volume thanks to the high-fidelity of 3G observations.

\begin{figure*}
\includegraphics[width=0.98\textwidth]{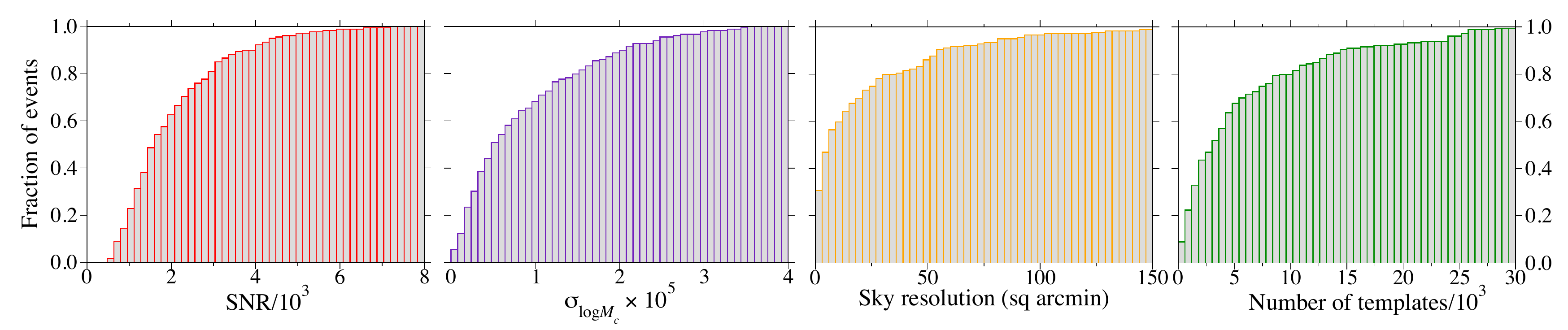}
\caption{The figure shows histograms of several parameters for the best 95\% of the events
that are visible both in 3G ground-based detectors and LISA (with an SNR $\ge 4$):
(i) the SNR, (ii) the error in the chirpmass, (iii) the error in 
the sky position, all determined by the 3G network, and (iv) the number of templates 
required to search for a stellar-mass BBH in the LISA data.}
\label{fig:num templates}
\end{figure*}

In one year, the ground-based 3G network of CE and ET would find ${\sim}200$ events 
that would have an SNR of 4 or more in LISA. For each such event we estimated 
the number of templates necessary for an archival search in LISA data. To this end, we 
computed the Fisher information matrix and its inverse---the variance-covariance matrix---to 
obtain the expected error in the measurement of various extrinsic and intrinsic
parameters using a 3G network of two CE detectors, one each in the USA and 
Australia, and one ET in Europe. Fig.\,\ref{fig:num templates} plots a subset of our
results: the distribution of the SNR, the error in the measurement of the chirpmass, 
and the uncertainty in the sky position (the left three panels). In making these plots
we have rejected the worst 5\% of the outliers in our simulation.

Due to the large SNR in the 3G network the parameters of the events are constrained 
very tightly. We assume that the events are contained within the 2-sigma region of the 
parameter uncertainties. This means that 90\% of the sources 
will be resolved by the 3G network to within $\sim 60$ arcmin$^2,$ which is better than 
the angular resolution of LISA for these sources; at an SNR of $\rho=10$, 
$\Delta\Omega_{\rm LISA} \sim [1.22 \lambda/(D\rho)]^2\sim 1800\,\rm arcmin^2,$ 
where $D=2\,\rm AU$ is LISA's baseline and $\lambda\sim3\times 10^7\,\rm m$ is the wavelength of GWs
corresponding to a frequency of 10 mHz.  
Moreover, the angles describing the orientation of the detector are also precisely 
determined by the 3G network---errors in the inclination and polarization angles are measured 
to within 1.0$^\circ$ and 2.3$^\circ$, respectively, for 90\% of the events.  Consequently, it is not 
necessary to include the angular parameters in the search nor the trigger time, which will be known
to better than 15 $\mu$s for 90\% of the events. For the Gaussian distribution of spins considered
in this paper it is not necessary to include spins either, leaving just the two masses over 
which a search should be carried out.

The number of templates required for an archival search for stellar-mass BBHs in the 
\emph{LISA band} for the ${\sim}200$ events in our population is shown in the right most 
panel of Fig.\,\ref{fig:num templates}. These numbers are computed using a minimal match 
of 0.95 (or allowing for a loss of less than 5\% in the SNR) \cite{Owen:1998dk}
and assuming that the true event lies in the 2-sigma region of the uncertainty in the masses
determined by the 3G network.  A vast majority (90\%) of the events require $< 15,000$ templates. 
One can employ singular value decomposition to find the number of \emph{independent} templates 
\cite{Cannon:2011vi}, which is typically 1 to 2 orders-of-magnitude smaller than the 
number of templates found at this minimal match, or ${\sim}150$-$1500$ for most of the events. 
Thus, an SNR-4 event in LISA will have a p-value $10^{-2}$ or smaller.

\paragraph*{Multiparameter tests from the higher PN side:}
One may consider an interesting variant of the multiparameter tests of GR where more than one PN parameter is treated as independent, starting from the highest order that is currently known, which is 3.5PN. This may be thought of as tests of alternatives to GR where up to a particular PN order, the predictions of both GR and its alternative match but beyond that they deviate. This may be naturally motivated from an effective field theoretic perspective where the deviations may appear when the binary dynamics proceeds beyond a certain scale of velocity or field strength~\cite{Endlich:2017tqa,Sennett:2019bpc}. 

\begin{figure}[hb]
\includegraphics[scale=0.35]{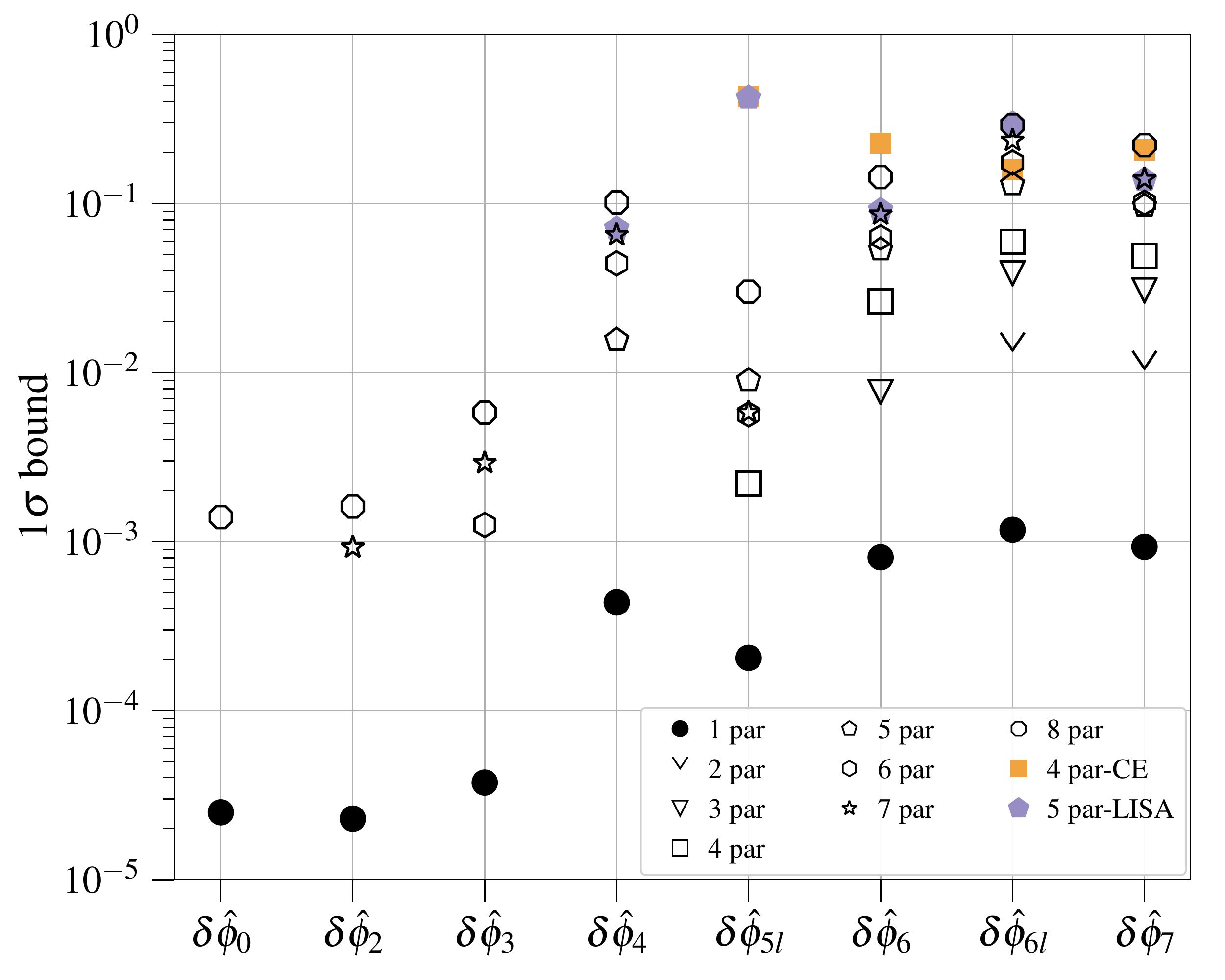}
	\caption{Multiparameter tests using multiband observations with LISA and CE. 
         Shown are combined $1\sigma$ bounds on various PN coefficients starting 
         from 3.5PN through 0PN in the inspiral phase of the signal while measuring 
         many of them together at a time. 
	  Different types of markers symbolize how 
         many PN deformation parameters were constrained simultaneously. 
        For example, `$\CIRCLE$'  represents `one PN deformation parameter at a time’, 
         `$\vee$' represents `two PN deformation parameters at a time’, and so on.
         The filled diamonds and pentagons are bounds obtained with CE and LISA,
         respectively, on the last four and five PN coefficients from their 
         respective golden binaries.  The total masses of the CE 
         and LISA golden binaries are $200 \,M_{\odot}$ 
         and $6.6\times10^{5}\,M_{\odot}$, both binaries are 1 Gpc away and have 
         component spins $\chi_1=0.6, \chi_2=0.5$.
         The figure represents results for the BBH population having Gaussian spin 
	 distribution (we get similar estimates for a uniform spin distribution).
	 }\label{fig:boundsright}
\end{figure}

Figure~\ref{fig:boundsright} shows the 
results for multiparameter tests starting from 3.5PN through 0PN order.
 For instance a three parameter test, would correspond to the case where only the last three PN deformation parameters (3.5PN, 3PN {\it log} and 3PN) are simultaneously estimated and so on.
The most significant result here is for the 7-parameter test for
which it is found that  the last seven PN parameters  can be bounded with
$\lesssim 10\%$ using the population we simulated earlier. 
This means, if we assume that the leading Newtonian 
%and next-to-leading phasing coefficients (0PN and 1PN) are not modified, 
coefficient is not modified, the simultaneous constraints on the remaining
seven are of the order of a few percent. Since a modification to the 
Newtonian phasing  would be at odds with the extremely stringent
bounds on them from binary pulsar observations~\cite{Yunes:2010qb}, this is the most general test we wish to carry out from this perspective.
Fig.~\ref{fig:boundsright} also shows the best bounds that CE  and LISA alone could yield from their golden binaries.
This demonstrates that multibanding of GW signals 
is the only way to put meaningful constraints on multiple PN deformation parameters with higher accuracy.

\paragraph*{Inversion Accuracy of the Fisher matrix:}
Due to large degeneracies among the PN deformation parameters, there are high chances that the
Fisher matrices (obtained from different events in the simulated population) corresponding to different multiparameter tests will be rendered ill-conditioned. 
Such Fisher matrices when inverted might lead to unreliable bounds. We therefore impose an inversion accuracy criterion on the Fisher matrices 
for all multiparameter tests. This criterion is defined to be $|\Gamma\cdot\Sigma -{\rm I}|\leq {\cal
O}(10^{-3})$, where $\Gamma$, $\Sigma$, and ${\rm I}$ are the multiband multiparameter Fisher matrix, the
corresponding variance-covariance matrix and the identity matrix, respectively. Any Fisher matrix obtained from an event corresponding a 
particular multi-parameter test, that does not meet this criterion is dropped from our analysis and is not used to obtain the combined multiband multiparameter bounds.

\bibliographystyle{apsrev}
\bibliography{Ref_list}

\begin{thebibliography}{52}
\expandafter\ifx\csname natexlab\endcsname\relax\def\natexlab#1{#1}\fi
\expandafter\ifx\csname bibnamefont\endcsname\relax
  \def\bibnamefont#1{#1}\fi
\expandafter\ifx\csname bibfnamefont\endcsname\relax
  \def\bibfnamefont#1{#1}\fi
\expandafter\ifx\csname citenamefont\endcsname\relax
  \def\citenamefont#1{#1}\fi
\expandafter\ifx\csname url\endcsname\relax
  \def\url#1{\texttt{#1}}\fi
\expandafter\ifx\csname urlprefix\endcsname\relax\def\urlprefix{URL }\fi
\providecommand{\bibinfo}[2]{#2}
\providecommand{\eprint}[2][]{\url{#2}}

\bibitem[{\citenamefont{Abbott et~al.}(2016{\natexlab{a}})}]{Discovery}
\bibinfo{author}{\bibfnamefont{B.~P.} \bibnamefont{Abbott}}
  \bibnamefont{et~al.} (\bibinfo{collaboration}{Virgo, LIGO Scientific}),
  \bibinfo{journal}{Phys. Rev. Lett.} \textbf{\bibinfo{volume}{116}},
  \bibinfo{pages}{061102} (\bibinfo{year}{2016}{\natexlab{a}}),
  \eprint{1602.03837}.

\bibitem[{\citenamefont{Abbott et~al.}(2019{\natexlab{a}})}]{GWTC1}
\bibinfo{author}{\bibfnamefont{B.~P.} \bibnamefont{Abbott}}
  \bibnamefont{et~al.} (\bibinfo{collaboration}{LIGO Scientific, Virgo}),
  \bibinfo{journal}{Phys. Rev.} \textbf{\bibinfo{volume}{X9}},
  \bibinfo{pages}{031040} (\bibinfo{year}{2019}{\natexlab{a}}),
  \eprint{1811.12907}.

\bibitem[{\citenamefont{Abbott et~al.}(2016{\natexlab{b}})}]{TOG}
\bibinfo{author}{\bibfnamefont{B.~P.} \bibnamefont{Abbott}}
  \bibnamefont{et~al.} (\bibinfo{collaboration}{Virgo, LIGO Scientific}),
  \bibinfo{journal}{Phys. Rev. Lett.} \textbf{\bibinfo{volume}{116}},
  \bibinfo{pages}{221101} (\bibinfo{year}{2016}{\natexlab{b}}),
  \eprint{1602.03841}.

\bibitem[{\citenamefont{Abbott et~al.}(2019{\natexlab{b}})}]{GWTC-TGR}
\bibinfo{author}{\bibfnamefont{B.~P.} \bibnamefont{Abbott}}
  \bibnamefont{et~al.} (\bibinfo{collaboration}{LIGO Scientific, Virgo}),
  \bibinfo{journal}{Phys. Rev.} \textbf{\bibinfo{volume}{D100}},
  \bibinfo{pages}{104036} (\bibinfo{year}{2019}{\natexlab{b}}),
  \eprint{1903.04467}.

\bibitem[{\citenamefont{Arun et~al.}(2006{\natexlab{a}})\citenamefont{Arun,
  Iyer, Qusailah, and Sathyaprakash}}]{AIQS06b}
\bibinfo{author}{\bibfnamefont{K.~G.} \bibnamefont{Arun}},
  \bibinfo{author}{\bibfnamefont{B.~R.} \bibnamefont{Iyer}},
  \bibinfo{author}{\bibfnamefont{M.~S.~S.} \bibnamefont{Qusailah}},
  \bibnamefont{and} \bibinfo{author}{\bibfnamefont{B.~S.}
  \bibnamefont{Sathyaprakash}}, \bibinfo{journal}{Phys.~Rev.~D}
  \textbf{\bibinfo{volume}{74}}, \bibinfo{pages}{024006}
  (\bibinfo{year}{2006}{\natexlab{a}}), \eprint{gr-qc/0604067}.

\bibitem[{\citenamefont{Mishra et~al.}(2010)\citenamefont{Mishra, Arun, Iyer,
  and Sathyaprakash}}]{MAIS10}
\bibinfo{author}{\bibfnamefont{C.~K.} \bibnamefont{Mishra}},
  \bibinfo{author}{\bibfnamefont{K.~G.} \bibnamefont{Arun}},
  \bibinfo{author}{\bibfnamefont{B.~R.} \bibnamefont{Iyer}}, \bibnamefont{and}
  \bibinfo{author}{\bibfnamefont{B.~S.} \bibnamefont{Sathyaprakash}},
  \bibinfo{journal}{Phys. Rev. {\bf D}} \textbf{\bibinfo{volume}{82}},
  \bibinfo{pages}{064010} (\bibinfo{year}{2010}), \eprint{1005.0304}.

\bibitem[{\citenamefont{Yunes and Pretorius}(2009)}]{YunesPretorius09}
\bibinfo{author}{\bibfnamefont{N.}~\bibnamefont{Yunes}} \bibnamefont{and}
  \bibinfo{author}{\bibfnamefont{F.}~\bibnamefont{Pretorius}},
  \bibinfo{journal}{Phys. Rev. D} \textbf{\bibinfo{volume}{80}},
  \bibinfo{pages}{122003} (\bibinfo{year}{2009}), \eprint{0909.3328}.

\bibitem[{\citenamefont{Li et~al.}(2012)}]{LiEtal2011}
\bibinfo{author}{\bibfnamefont{T.~G.~F.} \bibnamefont{Li}}
  \bibnamefont{et~al.}, \bibinfo{journal}{Phys.Rev. D}
  \textbf{\bibinfo{volume}{85}}, \bibinfo{pages}{082003}
  (\bibinfo{year}{2012}), \eprint{1111.5274}.

\bibitem[{\citenamefont{Blanchet}(2006)}]{Bliving}
\bibinfo{author}{\bibfnamefont{L.}~\bibnamefont{Blanchet}},
  \bibinfo{journal}{Living Rev. Rel.} \textbf{\bibinfo{volume}{9}},
  \bibinfo{pages}{4} (\bibinfo{year}{2006}), \eprint{arXiv:1310.1528}.

\bibitem[{\citenamefont{Sathyaprakash and
  Dhurandhar}(1991)}]{Sathyaprakash:1991mt}
\bibinfo{author}{\bibfnamefont{B.}~\bibnamefont{Sathyaprakash}}
  \bibnamefont{and}
  \bibinfo{author}{\bibfnamefont{S.}~\bibnamefont{Dhurandhar}},
  \bibinfo{journal}{Phys.\ Rev.\ D} \textbf{\bibinfo{volume}{44}},
  \bibinfo{pages}{3819} (\bibinfo{year}{1991}).

\bibitem[{\citenamefont{Blanchet et~al.}(1995)\citenamefont{Blanchet, Damour,
  Iyer, Will, and Wiseman}}]{BDIWW95}
\bibinfo{author}{\bibfnamefont{L.}~\bibnamefont{Blanchet}},
  \bibinfo{author}{\bibfnamefont{T.}~\bibnamefont{Damour}},
  \bibinfo{author}{\bibfnamefont{B.~R.} \bibnamefont{Iyer}},
  \bibinfo{author}{\bibfnamefont{C.~M.} \bibnamefont{Will}}, \bibnamefont{and}
  \bibinfo{author}{\bibfnamefont{A.~G.} \bibnamefont{Wiseman}},
  \bibinfo{journal}{Phys. Rev. Lett.} \textbf{\bibinfo{volume}{74}},
  \bibinfo{pages}{3515} (\bibinfo{year}{1995}), \eprint{gr-qc/9501027}.

\bibitem[{\citenamefont{Blanchet et~al.}(2004)\citenamefont{Blanchet, Damour,
  Esposito-Far{\`e}se, and Iyer}}]{BDEI04}
\bibinfo{author}{\bibfnamefont{L.}~\bibnamefont{Blanchet}},
  \bibinfo{author}{\bibfnamefont{T.}~\bibnamefont{Damour}},
  \bibinfo{author}{\bibfnamefont{G.}~\bibnamefont{Esposito-Far{\`e}se}},
  \bibnamefont{and} \bibinfo{author}{\bibfnamefont{B.~R.} \bibnamefont{Iyer}},
  \bibinfo{journal}{Phys. Rev. Lett.} \textbf{\bibinfo{volume}{93}},
  \bibinfo{pages}{091101} (\bibinfo{year}{2004}), \eprint{gr-qc/0406012}.

\bibitem[{\citenamefont{Arun et~al.}(2009)\citenamefont{Arun, Buonanno, Faye,
  and Ochsner}}]{ABFO08}
\bibinfo{author}{\bibfnamefont{K.~G.} \bibnamefont{Arun}},
  \bibinfo{author}{\bibfnamefont{A.}~\bibnamefont{Buonanno}},
  \bibinfo{author}{\bibfnamefont{G.}~\bibnamefont{Faye}}, \bibnamefont{and}
  \bibinfo{author}{\bibfnamefont{E.}~\bibnamefont{Ochsner}},
  \bibinfo{journal}{Phys. Rev. D} \textbf{\bibinfo{volume}{79}},
  \bibinfo{pages}{104023} (\bibinfo{year}{2009}), \eprint{0810.5336}.

\bibitem[{\citenamefont{Mishra et~al.}(2016)\citenamefont{Mishra, Kela, Arun,
  and Faye}}]{MKAF16}
\bibinfo{author}{\bibfnamefont{C.~K.} \bibnamefont{Mishra}},
  \bibinfo{author}{\bibfnamefont{A.}~\bibnamefont{Kela}},
  \bibinfo{author}{\bibfnamefont{K.~G.} \bibnamefont{Arun}}, \bibnamefont{and}
  \bibinfo{author}{\bibfnamefont{G.}~\bibnamefont{Faye}},
  \bibinfo{journal}{Phys. Rev.} \textbf{\bibinfo{volume}{D93}},
  \bibinfo{pages}{084054} (\bibinfo{year}{2016}), \eprint{1601.05588}.

\bibitem[{\citenamefont{Blanchet and Damour}(1988)}]{BD87}
\bibinfo{author}{\bibfnamefont{L.}~\bibnamefont{Blanchet}} \bibnamefont{and}
  \bibinfo{author}{\bibfnamefont{T.}~\bibnamefont{Damour}},
  \bibinfo{journal}{Phys. Rev.} \textbf{\bibinfo{volume}{D37}},
  \bibinfo{pages}{1410} (\bibinfo{year}{1988}).

\bibitem[{\citenamefont{Kidder et~al.}(1993)\citenamefont{Kidder, Will, and
  Wiseman}}]{KWWi93}
\bibinfo{author}{\bibfnamefont{L.}~\bibnamefont{Kidder}},
  \bibinfo{author}{\bibfnamefont{C.}~\bibnamefont{Will}}, \bibnamefont{and}
  \bibinfo{author}{\bibfnamefont{A.}~\bibnamefont{Wiseman}},
  \bibinfo{journal}{Phys. Rev. D} \textbf{\bibinfo{volume}{47}},
  \bibinfo{pages}{R4183} (\bibinfo{year}{1993}).

\bibitem[{\citenamefont{Blanchet et~al.}(2006)\citenamefont{Blanchet, Buonanno,
  and Faye}}]{BBuF06}
\bibinfo{author}{\bibfnamefont{L.}~\bibnamefont{Blanchet}},
  \bibinfo{author}{\bibfnamefont{A.}~\bibnamefont{Buonanno}}, \bibnamefont{and}
  \bibinfo{author}{\bibfnamefont{G.}~\bibnamefont{Faye}},
  \bibinfo{journal}{Phys. Rev.~D} \textbf{\bibinfo{volume}{{\bf 74}}},
  \bibinfo{pages}{104034} (\bibinfo{year}{2006}), \bibinfo{note}{erratum-ibid.D
  {\bf 75}, 049903 (E) (2007)}, \eprint{gr-qc/0605140}.

\bibitem[{\citenamefont{Buonanno et~al.}(2013)\citenamefont{Buonanno, Faye, and
  Hinderer}}]{BFH2012}
\bibinfo{author}{\bibfnamefont{A.}~\bibnamefont{Buonanno}},
  \bibinfo{author}{\bibfnamefont{G.}~\bibnamefont{Faye}}, \bibnamefont{and}
  \bibinfo{author}{\bibfnamefont{T.}~\bibnamefont{Hinderer}},
  \bibinfo{journal}{Phys.Rev.} \textbf{\bibinfo{volume}{D87}},
  \bibinfo{pages}{044009} (\bibinfo{year}{2013}), \eprint{1209.6349}.

\bibitem[{\citenamefont{Tagoshi et~al.}(1997)\citenamefont{Tagoshi, Mano, and
  Takasugi}}]{Tagoshi:1997jy}
\bibinfo{author}{\bibfnamefont{H.}~\bibnamefont{Tagoshi}},
  \bibinfo{author}{\bibfnamefont{S.}~\bibnamefont{Mano}}, \bibnamefont{and}
  \bibinfo{author}{\bibfnamefont{E.}~\bibnamefont{Takasugi}},
  \bibinfo{journal}{Prog. Theor. Phys.} \textbf{\bibinfo{volume}{98}},
  \bibinfo{pages}{829} (\bibinfo{year}{1997}), \eprint{gr-qc/9711072}.

\bibitem[{\citenamefont{Yunes et~al.}(2016)\citenamefont{Yunes, Yagi, and
  Pretorius}}]{YYP2016}
\bibinfo{author}{\bibfnamefont{N.}~\bibnamefont{Yunes}},
  \bibinfo{author}{\bibfnamefont{K.}~\bibnamefont{Yagi}}, \bibnamefont{and}
  \bibinfo{author}{\bibfnamefont{F.}~\bibnamefont{Pretorius}},
  \bibinfo{journal}{Phys. Rev.} \textbf{\bibinfo{volume}{D94}},
  \bibinfo{pages}{084002} (\bibinfo{year}{2016}), \eprint{1603.08955}.

\bibitem[{\citenamefont{Buonanno et~al.}(2009)\citenamefont{Buonanno, Iyer,
  Ochsner, Pan, and Sathyaprakash}}]{Buonanno:2009zt}
\bibinfo{author}{\bibfnamefont{A.}~\bibnamefont{Buonanno}},
  \bibinfo{author}{\bibfnamefont{B.}~\bibnamefont{Iyer}},
  \bibinfo{author}{\bibfnamefont{E.}~\bibnamefont{Ochsner}},
  \bibinfo{author}{\bibfnamefont{Y.}~\bibnamefont{Pan}}, \bibnamefont{and}
  \bibinfo{author}{\bibfnamefont{B.}~\bibnamefont{Sathyaprakash}},
  \bibinfo{journal}{Phys.\ Rev.\ D} \textbf{\bibinfo{volume}{80}},
  \bibinfo{pages}{084043} (\bibinfo{year}{2009}), \eprint{0907.0700}.

\bibitem[{\citenamefont{Arun et~al.}(2006{\natexlab{b}})\citenamefont{Arun,
  Iyer, Qusailah, and Sathyaprakash}}]{AIQS06a}
\bibinfo{author}{\bibfnamefont{K.~G.} \bibnamefont{Arun}},
  \bibinfo{author}{\bibfnamefont{B.~R.} \bibnamefont{Iyer}},
  \bibinfo{author}{\bibfnamefont{M.~S.~S.} \bibnamefont{Qusailah}},
  \bibnamefont{and} \bibinfo{author}{\bibfnamefont{B.~S.}
  \bibnamefont{Sathyaprakash}}, \bibinfo{journal}{Class. Quantum Grav.}
  \textbf{\bibinfo{volume}{23}}, \bibinfo{pages}{L37}
  (\bibinfo{year}{2006}{\natexlab{b}}), \eprint{gr-qc/0604018}.

\bibitem[{\citenamefont{Endlich et~al.}(2017)\citenamefont{Endlich, Gorbenko,
  Huang, and Senatore}}]{Endlich:2017tqa}
\bibinfo{author}{\bibfnamefont{S.}~\bibnamefont{Endlich}},
  \bibinfo{author}{\bibfnamefont{V.}~\bibnamefont{Gorbenko}},
  \bibinfo{author}{\bibfnamefont{J.}~\bibnamefont{Huang}}, \bibnamefont{and}
  \bibinfo{author}{\bibfnamefont{L.}~\bibnamefont{Senatore}},
  \bibinfo{journal}{JHEP} \textbf{\bibinfo{volume}{09}}, \bibinfo{pages}{122}
  (\bibinfo{year}{2017}), \eprint{1704.01590}.

\bibitem[{\citenamefont{Sennett et~al.}(2019)\citenamefont{Sennett, Brito,
  Buonanno, Gorbenko, and Senatore}}]{Sennett:2019bpc}
\bibinfo{author}{\bibfnamefont{N.}~\bibnamefont{Sennett}},
  \bibinfo{author}{\bibfnamefont{R.}~\bibnamefont{Brito}},
  \bibinfo{author}{\bibfnamefont{A.}~\bibnamefont{Buonanno}},
  \bibinfo{author}{\bibfnamefont{V.}~\bibnamefont{Gorbenko}}, \bibnamefont{and}
  \bibinfo{author}{\bibfnamefont{L.}~\bibnamefont{Senatore}}
  (\bibinfo{year}{2019}), \eprint{1912.09917}.

\bibitem[{\citenamefont{{Reitze} et~al.}(2019)}]{2019arXiv190704833R}
\bibinfo{author}{\bibfnamefont{D.}~\bibnamefont{{Reitze}}} \bibnamefont{et~al.}
  (\bibinfo{year}{2019}), \eprint{1907.04833}.

\bibitem[{\citenamefont{Punturo et~al.}(2010)}]{Punturo:2010zz}
\bibinfo{author}{\bibfnamefont{M.}~\bibnamefont{Punturo}} \bibnamefont{et~al.},
  \bibinfo{journal}{Class. Quant. Grav.} \textbf{\bibinfo{volume}{27}},
  \bibinfo{pages}{194002} (\bibinfo{year}{2010}).

\bibitem[{\citenamefont{{Amaro-Seoane}
  et~al.}(2017)\citenamefont{{Amaro-Seoane}, {Audley}, {Babak}, {Baker},
  {Barausse}, {Bender}, {Berti}, {Binetruy}, {Born}, {Bortoluzzi}
  et~al.}}]{LISA2017}
\bibinfo{author}{\bibfnamefont{P.}~\bibnamefont{{Amaro-Seoane}}},
  \bibinfo{author}{\bibfnamefont{H.}~\bibnamefont{{Audley}}},
  \bibinfo{author}{\bibfnamefont{S.}~\bibnamefont{{Babak}}},
  \bibinfo{author}{\bibfnamefont{J.}~\bibnamefont{{Baker}}},
  \bibinfo{author}{\bibfnamefont{E.}~\bibnamefont{{Barausse}}},
  \bibinfo{author}{\bibfnamefont{P.}~\bibnamefont{{Bender}}},
  \bibinfo{author}{\bibfnamefont{E.}~\bibnamefont{{Berti}}},
  \bibinfo{author}{\bibfnamefont{P.}~\bibnamefont{{Binetruy}}},
  \bibinfo{author}{\bibfnamefont{M.}~\bibnamefont{{Born}}},
  \bibinfo{author}{\bibfnamefont{D.}~\bibnamefont{{Bortoluzzi}}},
  \bibnamefont{et~al.}, \bibinfo{journal}{arXiv e-prints}
  \bibinfo{eid}{arXiv:1702.00786} (\bibinfo{year}{2017}), \eprint{1702.00786}.

\bibitem[{\citenamefont{Datta et~al.}(2020)\citenamefont{Datta, Gupta, Kastha,
  Arun, and Sathyaprakash}}]{Datta:2020vcj}
\bibinfo{author}{\bibfnamefont{S.}~\bibnamefont{Datta}},
  \bibinfo{author}{\bibfnamefont{A.}~\bibnamefont{Gupta}},
  \bibinfo{author}{\bibfnamefont{S.}~\bibnamefont{Kastha}},
  \bibinfo{author}{\bibfnamefont{K.}~\bibnamefont{Arun}}, \bibnamefont{and}
  \bibinfo{author}{\bibfnamefont{B.}~\bibnamefont{Sathyaprakash}}
  (\bibinfo{year}{2020}), \eprint{2006.12137}.

\bibitem[{\citenamefont{Nair et~al.}(2016)\citenamefont{Nair, Jhingan, and
  Tanaka}}]{Nair:2015bga}
\bibinfo{author}{\bibfnamefont{R.}~\bibnamefont{Nair}},
  \bibinfo{author}{\bibfnamefont{S.}~\bibnamefont{Jhingan}}, \bibnamefont{and}
  \bibinfo{author}{\bibfnamefont{T.}~\bibnamefont{Tanaka}},
  \bibinfo{journal}{PTEP} \textbf{\bibinfo{volume}{2016}},
  \bibinfo{pages}{053E01} (\bibinfo{year}{2016}), \eprint{1504.04108}.

\bibitem[{\citenamefont{Vitale}(2016)}]{Vitale:2016rfr}
\bibinfo{author}{\bibfnamefont{S.}~\bibnamefont{Vitale}},
  \bibinfo{journal}{Phys. Rev. Lett.} \textbf{\bibinfo{volume}{117}},
  \bibinfo{pages}{051102} (\bibinfo{year}{2016}), \eprint{1605.01037}.

\bibitem[{\citenamefont{Sesana}(2016)}]{Sesana:2016ljz}
\bibinfo{author}{\bibfnamefont{A.}~\bibnamefont{Sesana}},
  \bibinfo{journal}{Phys. Rev. Lett.} \textbf{\bibinfo{volume}{116}},
  \bibinfo{pages}{231102} (\bibinfo{year}{2016}), \eprint{1602.06951}.

\bibitem[{\citenamefont{Barausse et~al.}(2016)\citenamefont{Barausse, Yunes,
  and Chamberlain}}]{Barausse:2016eii}
\bibinfo{author}{\bibfnamefont{E.}~\bibnamefont{Barausse}},
  \bibinfo{author}{\bibfnamefont{N.}~\bibnamefont{Yunes}}, \bibnamefont{and}
  \bibinfo{author}{\bibfnamefont{K.}~\bibnamefont{Chamberlain}},
  \bibinfo{journal}{Phys. Rev. Lett.} \textbf{\bibinfo{volume}{116}},
  \bibinfo{pages}{241104} (\bibinfo{year}{2016}), \eprint{1603.04075}.

\bibitem[{\citenamefont{Toubiana et~al.}(2020)\citenamefont{Toubiana, Marsat,
  Babak, Barausse, and Baker}}]{Toubiana:2020vtf}
\bibinfo{author}{\bibfnamefont{A.}~\bibnamefont{Toubiana}},
  \bibinfo{author}{\bibfnamefont{S.}~\bibnamefont{Marsat}},
  \bibinfo{author}{\bibfnamefont{S.}~\bibnamefont{Babak}},
  \bibinfo{author}{\bibfnamefont{E.}~\bibnamefont{Barausse}}, \bibnamefont{and}
  \bibinfo{author}{\bibfnamefont{J.}~\bibnamefont{Baker}}
  (\bibinfo{year}{2020}), \eprint{2004.03626}.

\bibitem[{\citenamefont{Liu et~al.}(2020)\citenamefont{Liu, Shao, Zhao, and
  Gao}}]{Liu:2020nwz}
\bibinfo{author}{\bibfnamefont{C.}~\bibnamefont{Liu}},
  \bibinfo{author}{\bibfnamefont{L.}~\bibnamefont{Shao}},
  \bibinfo{author}{\bibfnamefont{J.}~\bibnamefont{Zhao}}, \bibnamefont{and}
  \bibinfo{author}{\bibfnamefont{Y.}~\bibnamefont{Gao}} (\bibinfo{year}{2020}),
  \eprint{2004.12096}.

\bibitem[{\citenamefont{Carson and Yagi}(2019)}]{Carson:2019rda}
\bibinfo{author}{\bibfnamefont{Z.}~\bibnamefont{Carson}} \bibnamefont{and}
  \bibinfo{author}{\bibfnamefont{K.}~\bibnamefont{Yagi}}
  (\bibinfo{year}{2019}), \eprint{1905.13155}.

\bibitem[{\citenamefont{Gnocchi et~al.}(2019)\citenamefont{Gnocchi, Maselli,
  Abdelsalhin, Giacobbo, and Mapelli}}]{Gnocchi:2019jzp}
\bibinfo{author}{\bibfnamefont{G.}~\bibnamefont{Gnocchi}},
  \bibinfo{author}{\bibfnamefont{A.}~\bibnamefont{Maselli}},
  \bibinfo{author}{\bibfnamefont{T.}~\bibnamefont{Abdelsalhin}},
  \bibinfo{author}{\bibfnamefont{N.}~\bibnamefont{Giacobbo}}, \bibnamefont{and}
  \bibinfo{author}{\bibfnamefont{M.}~\bibnamefont{Mapelli}},
  \bibinfo{journal}{Phys. Rev.} \textbf{\bibinfo{volume}{D100}},
  \bibinfo{pages}{064024} (\bibinfo{year}{2019}), \eprint{1905.13460}.

\bibitem[{\citenamefont{Abbott et~al.}(2016{\natexlab{c}})\citenamefont{Abbott,
  Abbott, Abbott, Abernathy, Acernese, Ackley, Adams, Adams, Addesso, Adhikari
  et~al.}}]{GW150914}
\bibinfo{author}{\bibfnamefont{B.}~\bibnamefont{Abbott}},
  \bibinfo{author}{\bibfnamefont{R.}~\bibnamefont{Abbott}},
  \bibinfo{author}{\bibfnamefont{T.}~\bibnamefont{Abbott}},
  \bibinfo{author}{\bibfnamefont{M.}~\bibnamefont{Abernathy}},
  \bibinfo{author}{\bibfnamefont{F.}~\bibnamefont{Acernese}},
  \bibinfo{author}{\bibfnamefont{K.}~\bibnamefont{Ackley}},
  \bibinfo{author}{\bibfnamefont{C.}~\bibnamefont{Adams}},
  \bibinfo{author}{\bibfnamefont{T.}~\bibnamefont{Adams}},
  \bibinfo{author}{\bibfnamefont{P.}~\bibnamefont{Addesso}},
  \bibinfo{author}{\bibfnamefont{R.}~\bibnamefont{Adhikari}},
  \bibnamefont{et~al.}, \bibinfo{journal}{Physical Review Letters}
  \textbf{\bibinfo{volume}{116}} (\bibinfo{year}{2016}{\natexlab{c}}), ISSN
  \bibinfo{issn}{1079-7114},
  \urlprefix\url{http://dx.doi.org/10.1103/PhysRevLett.116.061102}.

\bibitem[{\citenamefont{Baibhav et~al.}(2019)\citenamefont{Baibhav, Berti,
  Gerosa, Mapelli, Giacobbo, Bouffanais, and Di~Carlo}}]{Baibhav:2019gxm}
\bibinfo{author}{\bibfnamefont{V.}~\bibnamefont{Baibhav}},
  \bibinfo{author}{\bibfnamefont{E.}~\bibnamefont{Berti}},
  \bibinfo{author}{\bibfnamefont{D.}~\bibnamefont{Gerosa}},
  \bibinfo{author}{\bibfnamefont{M.}~\bibnamefont{Mapelli}},
  \bibinfo{author}{\bibfnamefont{N.}~\bibnamefont{Giacobbo}},
  \bibinfo{author}{\bibfnamefont{Y.}~\bibnamefont{Bouffanais}},
  \bibnamefont{and} \bibinfo{author}{\bibfnamefont{U.~N.}
  \bibnamefont{Di~Carlo}}, \bibinfo{journal}{Phys. Rev.}
  \textbf{\bibinfo{volume}{D100}}, \bibinfo{pages}{064060}
  (\bibinfo{year}{2019}), \eprint{1906.04197}.

\bibitem[{\citenamefont{Abbott
  et~al.}(2019{\natexlab{c}})}]{LIGOScientific:2018jsj}
\bibinfo{author}{\bibfnamefont{B.~P.} \bibnamefont{Abbott}}
  \bibnamefont{et~al.} (\bibinfo{collaboration}{LIGO Scientific, Virgo}),
  \bibinfo{journal}{Astrophys. J.} \textbf{\bibinfo{volume}{882}},
  \bibinfo{pages}{L24} (\bibinfo{year}{2019}{\natexlab{c}}),
  \eprint{1811.12940}.

\bibitem[{\citenamefont{Wong et~al.}(2018)\citenamefont{Wong, Kovetz, Cutler,
  and Berti}}]{Wong:2018uwb}
\bibinfo{author}{\bibfnamefont{K.~W.~K.} \bibnamefont{Wong}},
  \bibinfo{author}{\bibfnamefont{E.~D.} \bibnamefont{Kovetz}},
  \bibinfo{author}{\bibfnamefont{C.}~\bibnamefont{Cutler}}, \bibnamefont{and}
  \bibinfo{author}{\bibfnamefont{E.}~\bibnamefont{Berti}},
  \bibinfo{journal}{Phys. Rev. Lett.} \textbf{\bibinfo{volume}{121}},
  \bibinfo{pages}{251102} (\bibinfo{year}{2018}), \eprint{1808.08247}.

\bibitem[{\citenamefont{Rao}(1945)}]{Rao45}
\bibinfo{author}{\bibfnamefont{C.}~\bibnamefont{Rao}},
  \bibinfo{journal}{Bullet. Calcutta Math. Soc} \textbf{\bibinfo{volume}{37}},
  \bibinfo{pages}{81} (\bibinfo{year}{1945}).

\bibitem[{\citenamefont{Cramer}(1946)}]{Cramer46}
\bibinfo{author}{\bibfnamefont{H.}~\bibnamefont{Cramer}},
  \emph{\bibinfo{title}{Mathematical methods in statistics}}
  (\bibinfo{publisher}{Pergamon Press}, \bibinfo{address}{Princeton University
  Press, NJ, U.S.A.}, \bibinfo{year}{1946}).

\bibitem[{\citenamefont{Cutler and Flanagan}(1994)}]{CF94}
\bibinfo{author}{\bibfnamefont{C.}~\bibnamefont{Cutler}} \bibnamefont{and}
  \bibinfo{author}{\bibfnamefont{E.}~\bibnamefont{Flanagan}},
  \bibinfo{journal}{Phys. Rev. D} \textbf{\bibinfo{volume}{49}},
  \bibinfo{pages}{2658} (\bibinfo{year}{1994}).

\bibitem[{\citenamefont{{Berti} et~al.}(2005)\citenamefont{{Berti}, {Buonanno},
  and {Will}}}]{BBW05a}
\bibinfo{author}{\bibfnamefont{E.}~\bibnamefont{{Berti}}},
  \bibinfo{author}{\bibfnamefont{A.}~\bibnamefont{{Buonanno}}},
  \bibnamefont{and} \bibinfo{author}{\bibfnamefont{C.~M.}
  \bibnamefont{{Will}}}, \bibinfo{journal}{Phys.~Rev.~D}
  \textbf{\bibinfo{volume}{71}}, \bibinfo{pages}{084025}
  (\bibinfo{year}{2005}), \eprint{gr-qc/0411129}.

\bibitem[{\citenamefont{Husa et~al.}(2016)\citenamefont{Husa, Khan, Hannam,
  P{\"u}rrer, Ohme, Forteza, and Boh{\'e}}}]{Husa2016}
\bibinfo{author}{\bibfnamefont{S.}~\bibnamefont{Husa}},
  \bibinfo{author}{\bibfnamefont{S.}~\bibnamefont{Khan}},
  \bibinfo{author}{\bibfnamefont{M.}~\bibnamefont{Hannam}},
  \bibinfo{author}{\bibfnamefont{M.}~\bibnamefont{P{\"u}rrer}},
  \bibinfo{author}{\bibfnamefont{F.}~\bibnamefont{Ohme}},
  \bibinfo{author}{\bibfnamefont{X.~J.} \bibnamefont{Forteza}},
  \bibnamefont{and} \bibinfo{author}{\bibfnamefont{A.}~\bibnamefont{Boh{\'e}}},
  \bibinfo{journal}{Physical Review D} \textbf{\bibinfo{volume}{93}}
  (\bibinfo{year}{2016}), ISSN \bibinfo{issn}{2470-0029},
  \urlprefix\url{http://dx.doi.org/10.1103/PhysRevD.93.044006}.

\bibitem[{\citenamefont{Khan et~al.}(2016)\citenamefont{Khan, Husa, Hannam,
  Ohme, P{\"u}rrer, Forteza, and Boh{\'e}}}]{Khan2016}
\bibinfo{author}{\bibfnamefont{S.}~\bibnamefont{Khan}},
  \bibinfo{author}{\bibfnamefont{S.}~\bibnamefont{Husa}},
  \bibinfo{author}{\bibfnamefont{M.}~\bibnamefont{Hannam}},
  \bibinfo{author}{\bibfnamefont{F.}~\bibnamefont{Ohme}},
  \bibinfo{author}{\bibfnamefont{M.}~\bibnamefont{P{\"u}rrer}},
  \bibinfo{author}{\bibfnamefont{X.~J.} \bibnamefont{Forteza}},
  \bibnamefont{and} \bibinfo{author}{\bibfnamefont{A.}~\bibnamefont{Boh{\'e}}},
  \bibinfo{journal}{Physical Review D} \textbf{\bibinfo{volume}{93}}
  (\bibinfo{year}{2016}), ISSN \bibinfo{issn}{2470-0029},
  \urlprefix\url{http://dx.doi.org/10.1103/PhysRevD.93.044007}.

\bibitem[{\citenamefont{Robson et~al.}(2019)\citenamefont{Robson, Cornish, and
  Liu}}]{Cornish:2018dyw}
\bibinfo{author}{\bibfnamefont{T.}~\bibnamefont{Robson}},
  \bibinfo{author}{\bibfnamefont{N.~J.} \bibnamefont{Cornish}},
  \bibnamefont{and} \bibinfo{author}{\bibfnamefont{C.}~\bibnamefont{Liu}},
  \bibinfo{journal}{Class. Quant. Grav.} \textbf{\bibinfo{volume}{36}},
  \bibinfo{pages}{105011} (\bibinfo{year}{2019}), \eprint{1803.01944}.

\bibitem[{\citenamefont{Poisson and Will}(1995)}]{PW95}
\bibinfo{author}{\bibfnamefont{E.}~\bibnamefont{Poisson}} \bibnamefont{and}
  \bibinfo{author}{\bibfnamefont{C.}~\bibnamefont{Will}},
  \bibinfo{journal}{Phys. Rev. D} \textbf{\bibinfo{volume}{52}},
  \bibinfo{pages}{848} (\bibinfo{year}{1995}).

\bibitem[{\citenamefont{Moore et~al.}(2019)\citenamefont{Moore, Gerosa, and
  Klein}}]{Moore:2019pke}
\bibinfo{author}{\bibfnamefont{C.~J.} \bibnamefont{Moore}},
  \bibinfo{author}{\bibfnamefont{D.}~\bibnamefont{Gerosa}}, \bibnamefont{and}
  \bibinfo{author}{\bibfnamefont{A.}~\bibnamefont{Klein}},
  \bibinfo{journal}{Mon. Not. Roy. Astron. Soc.}
  \textbf{\bibinfo{volume}{488}}, \bibinfo{pages}{L94} (\bibinfo{year}{2019}),
  \eprint{1905.11998}.

\bibitem[{\citenamefont{Owen and Sathyaprakash}(1999)}]{Owen:1998dk}
\bibinfo{author}{\bibfnamefont{B.~J.} \bibnamefont{Owen}} \bibnamefont{and}
  \bibinfo{author}{\bibfnamefont{B.}~\bibnamefont{Sathyaprakash}},
  \bibinfo{journal}{Phys. Rev. D} \textbf{\bibinfo{volume}{60}},
  \bibinfo{pages}{022002} (\bibinfo{year}{1999}), \eprint{gr-qc/9808076}.

\bibitem[{\citenamefont{Cannon et~al.}(2012)}]{Cannon:2011vi}
\bibinfo{author}{\bibfnamefont{K.}~\bibnamefont{Cannon}} \bibnamefont{et~al.},
  \bibinfo{journal}{Astrophys. J.} \textbf{\bibinfo{volume}{748}},
  \bibinfo{pages}{136} (\bibinfo{year}{2012}), \eprint{1107.2665}.

\bibitem[{\citenamefont{Yunes and Hughes}(2010)}]{Yunes:2010qb}
\bibinfo{author}{\bibfnamefont{N.}~\bibnamefont{Yunes}} \bibnamefont{and}
  \bibinfo{author}{\bibfnamefont{S.~A.} \bibnamefont{Hughes}},
  \bibinfo{journal}{Phys. Rev. D} \textbf{\bibinfo{volume}{82}},
  \bibinfo{pages}{082002} (\bibinfo{year}{2010}), \eprint{1007.1995}.

\end{thebibliography}

\end{document}